\PassOptionsToPackage{pdfpagelabels=true}{hyperref} 

\documentclass[fleqn,usenatbib]{mnras}

\usepackage{newtxtext,newtxmath}
\usepackage[utf8]{inputenc}

\usepackage[T1]{fontenc}
\usepackage{ae,aecompl}

\usepackage{graphicx}	
\usepackage{amsmath}	
\usepackage{newtxtext,newtxmath}
\usepackage{multirow}
\usepackage[table,xcdraw]{xcolor}

\title[]{KINETIC IMPACT AND GRAVITATIONAL PERTURBATIONS FOR ASTEROID DEFLECTION} 

\author[Chagas, B. S. et al.]{
Bruno Chagas,$^{1}$\thanks{E-mail: bruno.chagas@unesp.br}
Antonio F.B. de A. Prado,$^{1,2,3}$\thanks{E-mail: abertachinipr@hotmail.com}
Othon C. Winter,$^{1}$\thanks{E-mail: othon.winter@unesp.br} 
\\
$^{1}$UNESP, São Paulo State University, Grupo de Dinâmica Orbital e Planetologia, Guaratinguetá, CEP 12516-410, SP, Brazil\\
$^{2}$National Institute for Space Research, INPE, astronaut av., São José dos Campos-SP, 12227-010, SP, Brazil\\
$^{3}$Academy of Engineering, RUDN University, Miklukho-Maklaya Street 6, Moscow, 117198, Russia\\
}

\pubyear{2021}

\begin{document}
\label{firstpage}
\pagerange{\pageref{firstpage}--\pageref{lastpage}}
\maketitle

\begin{abstract}
Asteroids have called the attention of researchers around the world. Its chemical and physical composition can give us important information about the formation of our Solar System. In addition, the hypothesis of mining some of these objects is considered, since they contain precious metals. However, some asteroids have their orbits close to the orbit of the Earth. These nearby objects can pose a danger to our life in the planet, since some of them are large enough to cause catastrophic damage to the Earth. We will pay attention to the theme of deflecting a potentially dangerous asteroid. There are currently two main forms of this deviation: i) the impact of an object at high velocity with the asteroid, which can be a space vehicle or a smaller asteroid; ii) the use of a gravitational "tractor", which consist in placing an object (another asteroid or part of an asteroid), close to the body that is approaching the Earth, such that this gravitational interference can deflect its trajectory. In this work, we will evaluate the influence of gravitational perturbations in the most commonly mentioned asteroid deflection model in the literature, the kinetic impact deflection technique. With the impact, it is intended to change the kinetic energy of the asteroid, changing its orbit enough so that it does not present risks of impacts with the Earth.
\end{abstract}




\section{Introduction}

Asteroids have drawn the attention of researchers around the world for several reasons. Studies of the origin of the solar system, exploration and possible impacts with the Earth have been made. The complexity of deflecting asteroids with dangerous orbits to Earth due to their close approach has gained more and more attention and ideas for performing the deflection of an asteroid. Carusi et al. (2002)\cite{Carusi:2002dg} makes a study on the possibility of deflecting asteroids by means of a Delta v impulse in their orbit and found that the place where the impulse should be carried out must be considered, which already mentions the legitimacy of exploring the gravitational attraction when the asteroid has more than one close encounter with the planet before a possible impact.

Giving their contribution, Ahrens and Harris (1992) \cite{Ahrens:1992dg} also studied some forms of asteroid deflection, including the efficiency of fragmentation by nuclear deflection, among others, such as deflection by direct impact. Ledkov et al. (2014)\cite{Ledkov:2014dg} presented results for the proposal of using a smaller asteroid that is induced by a spacecraft using the gravitational tractor to redirect the small asteroid to impact a larger asteroid that will get closer to Earth. Fahnestockÿ and Scheeres (2008)\cite{Fahnestock:2008dg} also studied the gravitational tractor deflection technique, but to deflect the potentially dangerous object.

For Youtao and Jingyun (2016)\cite{Gao:2016dg}, the use of the technique of using a solar sail attached to the surface of the asteroid proved to be efficient. Negri et al. (2018)\cite{Negri:2018dg} studied the kinetic impact using Several values for the impulse and the influence of Jupiter on the disturbances involved. Izzo (2007)\cite{Izzo:2007dg} makes a comparison with the kinetic impact technique and the long-term impulse technique. It demonstrates the greater efficiency of the kinetic impact deviation technique. DJ Scheeres and RL Schweickart (2004)\cite{Scheeres:2004dg} carried out a series of studies related to maneuvers involving asteroid deviation, where the authors realistically debate the proposal of attaching a tug to an asteroid, where the efficiency and challenges involved in this proposal are analyzed.

Park and Mazanek (2003)\cite{Park:2003dg} mention of the importance of early warning in relation to discovering a potential impactor. Sanches et al. (2010)\cite{Sanchez:2010dg} presents a series of studies involving the solar collector, nuclear interceptor, kinetic impactor, low-thrust propulsion, mass driver, and gravity tractor. Cheng et al. (2015)\cite{Cheng:2015dg} demonstrate the objectives and what is expected from the DART mission, which will be the first asteroid diversion mission ever carried out, where a spacecraft will impact the smallest body of the binary system [65803] Didymos.

Sannikova and Kholshevnikov (2015)\cite{Sannikova:2015dg} present studies on a model where an object suffers the gravitational influence of a central body and a disturbing acceleration, showing that this model can be used in the deflection of asteroids. With this in mind, we used the work by Ferreira at al. (2017)\cite{Ferreira:2017dg}, which carried out a mapping of the energy variation involved in a propelled Swing By maneuver. We will use some of their analyzes to develop some topics found in our results. Our work will present the results of numerical simulations using the N-body problem, where we will analyze the influence of long-term gravitational perturbations. For this, we are considering the impulses to be applied at perihelion and aphelion of the orbit of the asteroid.

\section{Development} 
We are addressing the N-body problem in our work. As the dynamics involved in solving these problems are quite complex, we are carrying out numerical simulations.
We selected the asteroid Bennu, which was the target of the Osiris Rex mission and is part of the group of NEAs.

We used the Mercury integrator package, choosing the Burlish-Stoer integrator and, for the input data, we chose to use the Cartesian coordinates of bodies involved in our simulations. All the planets in the system solar, the moon and the asteroid were included in the simulations.

We do not consider the dimensional characteristics of the bodies, treating them as points of masses. We also do not consider other disturbing forces besides gravitational forces.

We found the input data from the JPL Horizons website on August 5, 2021. The values can be found in Tables 1 and 2, and, we started the simulations 100 years to find the closest asteroid-Earth approach. We found that it will be in 2080. We add the values of the position and velocity components found at this closest approximation and went back 100 years in time. The question is: "what happens in 100 years of perturbation after the impulses are applied to the orbit of the asteroid by the kinetic impact?"

It is relevant to clarify that we are only simulating the impulses through variations in the velocity of the asteroid at the beginning of the simulations, that is, we are not simulating the kinetic impact itself.

After going back in time for 100 years, we performed the simulations over time again, but now varying the velocity of the asteroid. The velocity variations used are in the range from -50 mm/s to 50 mm/s, with an increment of 10 mm/s. To improve our understanding of the results and better analyze the impulse locations, we initially divided the orbital period of the asteroid into 36 parts and went further and further back in time, applying the velocity variations at each point. We performed simulations separately at the perihelion and aphelion points.

\subsection{Mathematical Development}

The problem treated here refers to the problem of N-bodies, where we will be considering the planets of the solar system and the asteroid Bennu, which is the target of our study. We are considering the Sun as the origin of the system, so the equation of motion for the N-body system with a mass body much larger than the others can be written as (Dandy, 1992)\cite{Danby:1992dg}

\begin{equation}
\frac{d^2(\vec{r_i})}{dt^2} + G(m_n + m_i) \frac{\vec{r_i}}{r_{in} ^ 3}  
= G\sum_{j=1, j\neq i}^{N - 1}m_j\left(\frac{\vec{r_j} - \vec{r_i}}{r_{ij} ^ 3} - \frac{\vec{r_j}}{r_{jn} ^ 3}\right)
\end{equation}
where  $\vec{r_i}$ and $\vec{r_j}$  are the position vectors of particles i and j with respect to the central body, G the gravitational constant, $m_n$ the mass of the central body, $m_i$ and $m_j$ the masses of particles, $r_ {in}$ and $r_ { jn}$ are the relative distances between particles i and j with respect to the central body and $r_ {ij}$ the relative distance between particles.

We noticed that there are variations in the behavior of the semi-major axis of the asteroid over time due to the impulse applied. Equation 2 makes clear that, by varying the semi-major axis, we vary the energy of the asteroid.

\begin{equation}
 \label{eq:one}
 E = -\frac{\mu}{2a} 
\end{equation}
where $\mu_{M_2} = Gm_{M_2}$ and `a' is the semi-major axis of the asteroid.

Naturally, we also verify the evolution of the eccentricity over time and through the expression 3 we realize that the velocity varies proportionally with the eccentricity. This fact is very important for our final analyses.

\begin{equation}
 \label{eq:one}
 v = \sqrt{\frac{\mu}{p}}\left[esin(f)\textbf{$e_r$} + \left(1 + ecos(f)\right) \textbf{$e_s$}\right] 
\end{equation}
where e is the eccentricity, p is the semilatus rectum, given by p = $h^2$/$\mu$, where h is the orbital angular momentum, f is the true anomaly, $e_r$ is the unit vector in the radial direction and $e_s$ is the unit vector in the circumferential direction.

With the evolution of the research, we realized that our results had a great participation of the maneuvers assisted by gravity, `swing by', where this proved to be the main cause of change in the orbit of the asteroid. `Swing by' maneuvers have been known for a long time in the astronomical scene and therefore are not new. The model can be found in Prado (1996)\cite{Prado:1996dg} and Ferreira et al. (2017)\cite{Ferreira:2017dg}.

We can formulate the maneuver mathematically by

\begin{equation}
\Delta E = -2 V_2 V_{\infty} sin(\delta) sin(\psi)
\end{equation}
where $\Delta$ E is the variation of energy, $V_2$ is the velocity of $M_2$ with respect to $M_1$, $V_\infty$ is the particle velocity modulus at infinity before or after passing close to a body of mass m, $\delta$ is the deflection angle and represents half the rotation of the velocity vector due to the `swing by' and $\psi$ is the angle between the periapsis line and the
line $M_1${-}$M_2$, where,

\begin{equation}
sin(\delta) = \frac{1}{1 + \frac{r_p V_{\infty}^2}{\mu_2}} 
\end{equation}
$r_p$ is the minimum approximation distance between the bodies involved and $\mu_j$ = $Gm_j$ (Broucke, (1988))\cite{Broucke:1988dg}.

\begin{table}[h!]
\begin{center}
\caption{\label{tab:example}Coordinates of each body on August 5, 2021. Data taken from JPL NASA.}
\begin{tabular}{cccc}
\hline
\multicolumn{1}{|c|}{Bodies} & \multicolumn{1}{c|}{x(au)} & \multicolumn{1}{c|}{y(au)} & \multicolumn{1}{c|}{z(au)}\\
\hline
Mercury & $-2,896323130935082E-01$ & $1,850577763255397E-01$ & $4,169019903669655E-02$ \\
Venus & $-5,692473108639921E-01$ & $-4,448853077740749E-01$ & $2,674283485411453E-02$ \\
Earth & $ 6,855761418396654E-01$ & $-7,477764241622177E-01$ & $ 3,211115876042908E-05$ \\
Moon & $ 6,855266149744332E-01$ & $-7,451016724453864E-01$ & $ 1,230222158011698E-04$ \\
Mars & $-1,614905800650153E+00$ & $3,955532232885259E-01$ & $4,790302296698644E-02$ \\
Jupiter & $ 4,148571639394127E+00$ & $-2,841339576414095E+00$ & $-8,101578326784213E-02$ \\
Saturn & $ 6,390745065584204E+00$ & $-7,624741172250260E+00$ & $-1,217720468705376E-01$ \\
Uranus & $ 1,479576471988037E+01$ & $ 1,307144787474826E+01$ & $-1,431624026194872E-01$ \\
Neptune & $ 2,956816826343533E+01$ & $-4,557431754532644E+00$ & $-5,876364308100537E-01$ \\
Bennu & $-8,178122041042245E-01$ & $6,126618626692276E-01$ & $6,774426832917149E-02$ \\
\hline
\end{tabular}
\end{center}
\end{table}

\begin{table}[h!]
\centering
\caption{\label{tab:example}Velocity of each body on August 5, 2021. Data taken from JPL NASA.}
\begin{tabular}{cccc}
\hline
\multicolumn{1}{|c|}{Bodies} & \multicolumn{1}{c|}{vx(au/day)} & \multicolumn{1}{c|}{vy(au/day)} & \multicolumn{1}{c|}{vz(au/day)}\\
\hline
Mercury & $-2,090746004400052E-02$ & $-2,253446087272209E-02$ & $7,638025951147821E-05$ \\
Venus & $ 1,231235542811912E-02$ & $-1,603266518952336E-02$ & $-9,305216171721341E-04$ \\
Earth & $ 1,240809069382479E-02$ & $ 1,156282606438427E-02$ & $-1,150024366950303E-06$ \\
Moon & $ 1,184355116475734E-02$ & $ 1,153246162528540E-02$ & $ 4,458968850063871E-05$ \\
Mars & $-2,807629223103034E-03$ & $-1,239693158153517E-02$ & $-1,909313502062999E-04$ \\
Jupiter & $ 4,177326698222665E-03$ & $ 6,587451407751826E-03$ & $-1,208714010214003E-04$ \\
Saturn & $3.969863463512147E-03$ & $3.576887186643701E-03$ & $-2.197897896266317E-04$ \\
Uranus & $-2,628949090335184E-03$ & $2,770739109958900E-03$ & $4,429549807400130E-05$ \\
Neptune & $ 4,618822543692893E-04$ & $ 3,128329541906231E-03$ & $-7,516594581186683E-05$ \\
Bennu & $-1,309992480373342E-02$ & $-1,191701206869430E-02$ & $-1,210612917719958E-03$ \\
\hline
\end{tabular}
\par
\end{table}

\section{Results}\label{sec:results}

As mentioned before, we will focus on measuring the effects of the application of the impulses at perihelion and aphelion of the orbit of the asteroid. Figure 1 expresses the closest approximations in the interval of approximately 100 years in relation to the magnitude of each impulse applied. We will be able to see that we have achieved good results in terms of moving the asteroid away from the Earth, which is our priority in this work. We also see that by applying impulses in these two regions give better results. However, we will find some scenarios where the asteroid gets even closer to the Earth with the application of the impulse, as found in the impulse of 20 mm/s applied at perihelion and the impulses applied with magnitudes of -30 mm/s, 10 mm/s and 40 mm/s at aphelion.

\begin{figure*}[h]
    \begin{center}
        \includegraphics[width=0.95\textwidth, height=3.65in]{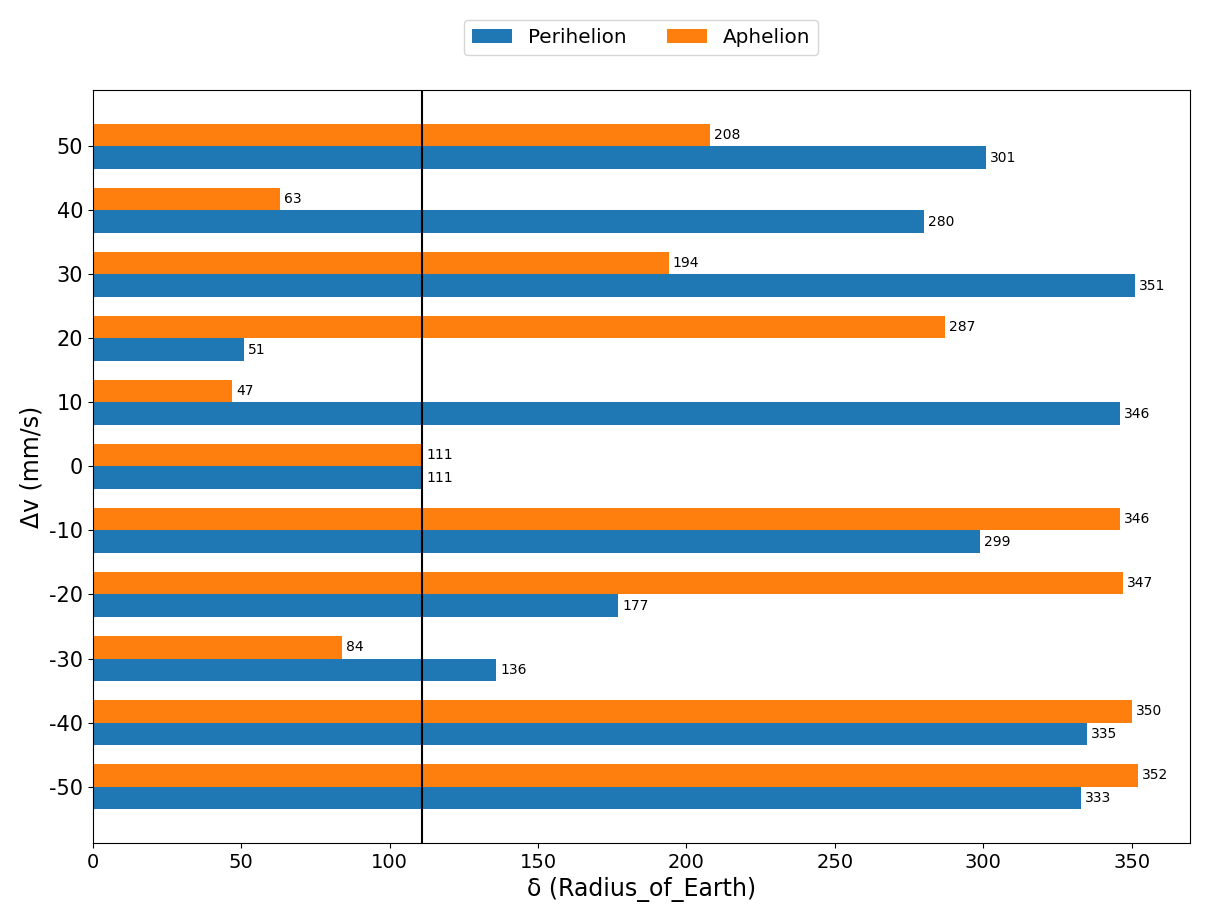}
        \caption{\small{Closer approximations between the asteroid and the Earth in relation to each impulse applied.}}
        \label{fig:fig1}
    \end{center}
\end{figure*}

It is interesting to note that the smaller impulses (-20mm/s, -10 mm/s, 10 mm/s and 20 mm/s) proved to be close or even as efficient as larger impulses, which is very encouraging when we think about the size of the impactor needed to make this impulse. Thinking about the size of the impactor, let us focus on the smallest impulses. Analyzing Figure 1, we will find an interesting scenario in the impulse of 10 mm/s, where the impulse applied at perihelion moves the asteroid considerably away from the Earth, while the same impulse applied at aphelion brings the asteroid closer to Earth.

Figure 2 shows the relative distances between the asteroid and the Earth, to better understand the evolution of the approaches between the bodies. Figure 2 (Top) shows the relative distance between the asteroid and Earth for the impulses applied at the perihelion of the orbit of the asteroid, where we find very interesting information. Initially, we can see that the asteroid has several close encounters with Earth during the simulation period. This encounter may be a little closer, since we are limiting the analysis to a relative distance of 2500 Earth Radius, where we have a conception that applying the impulse of 10 mm/s, the orbit of the asteroid changes, moving the asteroid away after the first major encounter before 20 years after the impulse. It is noticeable that there is a small variation in this region and that, analyzing each approach after this first major approach, it is noticed that the variations in these closer approaches intensify, which leads us to believe that the second closest approach to Earth already has a variation at greater relative distance. Those results will be useful for deeper analysis of the effects that we are demonstrating. It is also interesting to note that, for the 20 mm/s impulse to be the only impulse applied at perihelion that brought the asteroid even closer to Earth, this approach occurs at a different date and not at the end of the 100 years, occurring now between 70 and 80 years after the impulse is applied. It is good to understand that, although this approach occurs, it did not approach the asteroid at the level of a possible impact with Earth.

Figure 2 (lower) shows the evolution of the relative distances for the impulses applied at the aphelion of the orbit of the asteroid. Here, the variations in the relative distance in the second close encounter are more evident. We also see that, for the impulses that brought the asteroid even closer to Earth, these impulses occurred on different dates. 

\begin{figure*}[h!]
    \begin{center}
        \includegraphics[width=0.95\textwidth, height=3.65in]{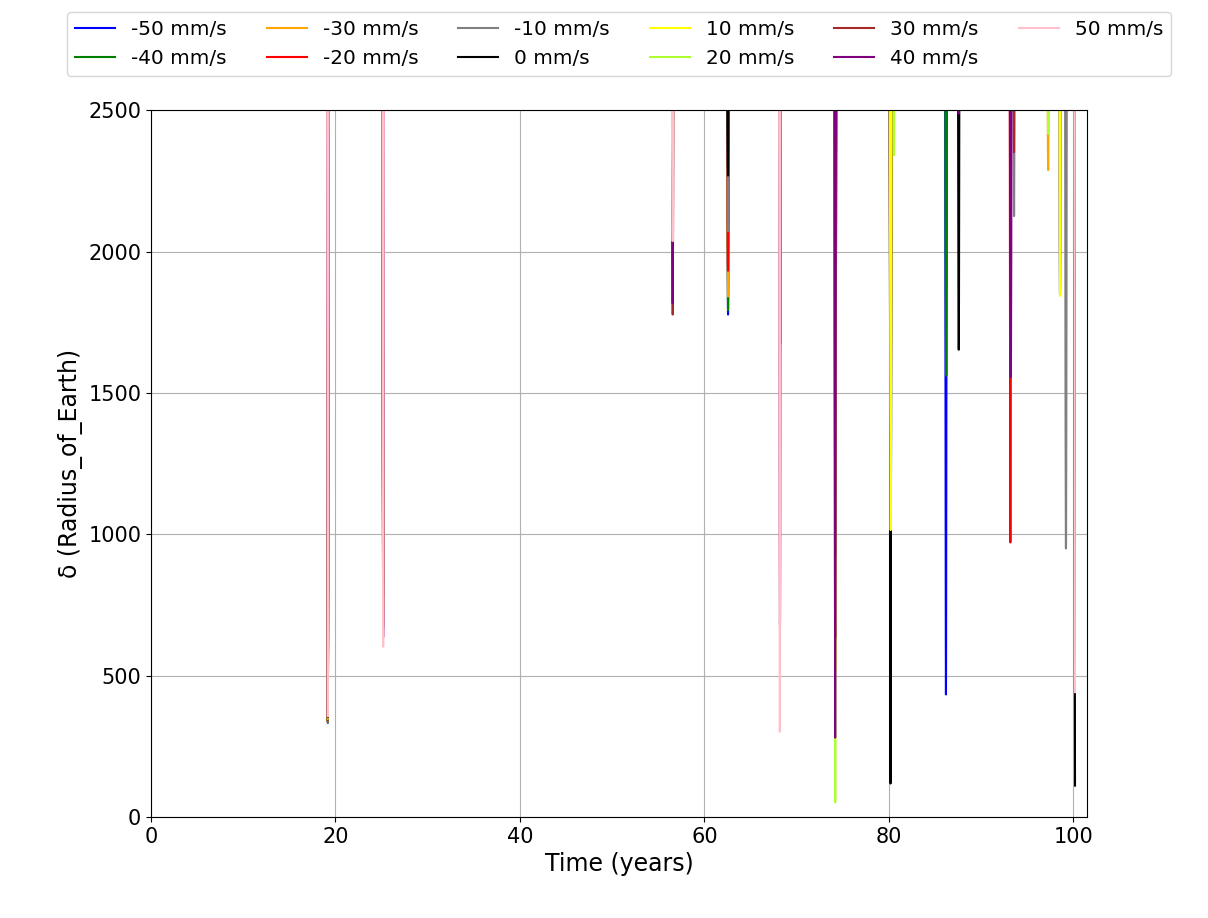}
        \par
        \includegraphics[width=0.95\textwidth, height=3.65in]{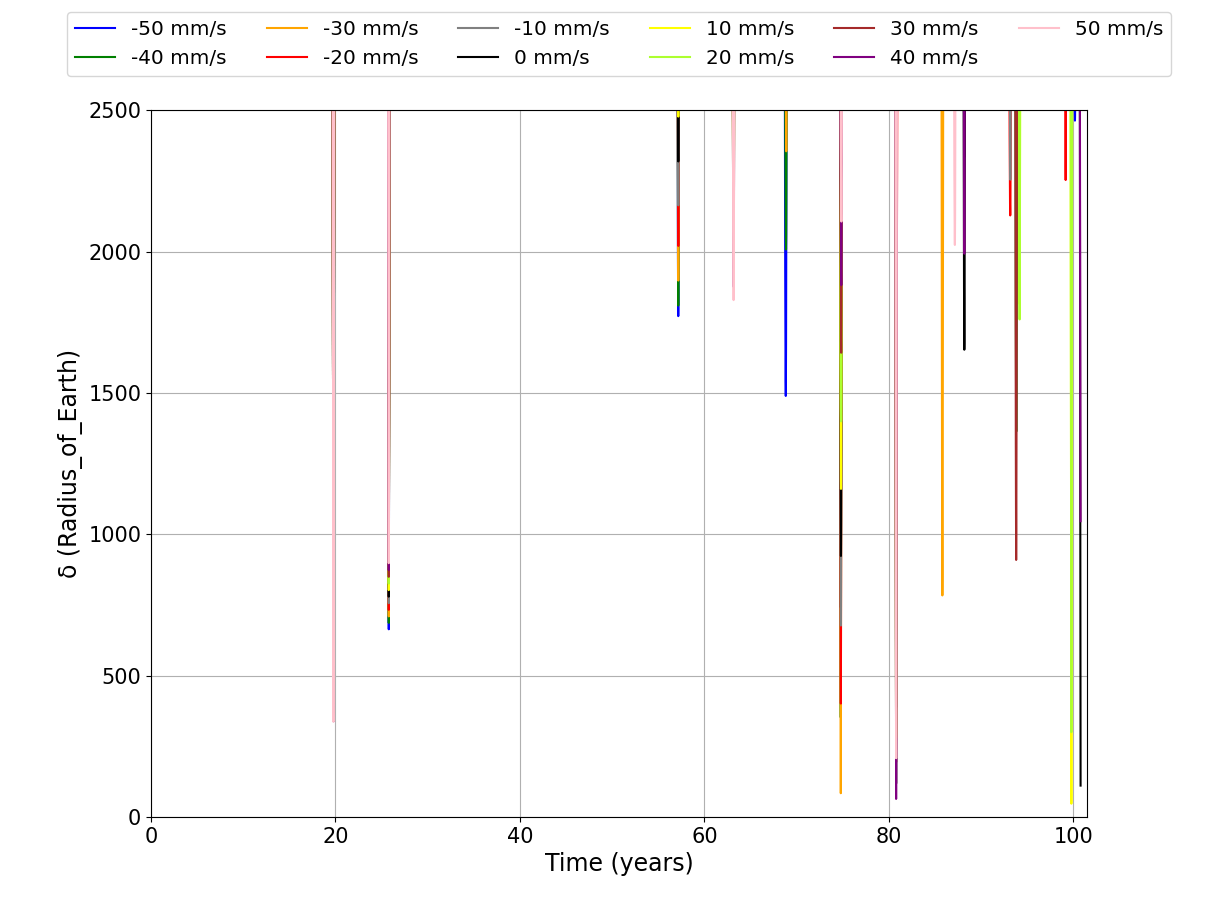}
        \caption{\small{Behavior of the relative distances between asteroid and Earth over time. The impulses are applied at perihelion (top), and aphelion (low)}}
        \label{fig:fig1}
    \end{center}
\end{figure*}

It is evident that the orbit of the asteroid is varying. We can find it through the behavior of the semi-major axis over time (Figure 3). Figure 3 (left) shows the evolution of the semi-major axis for impulses applied at perihelion and Figure 3 (right) shows the evolution of the semi-major axis for impulses applied at the aphelion of the orbit of the asteroid. Analyzing this figure, we will not find any indication if there is any pattern variation that can express whether the asteroid will approach or move away from the Earth. However, we can identify variations in the semi-major axis of the asteroid in its first approach which variations are intensified in the second close encounter. It intensifies until a new large close encounter occurs that causes even larger variations. What seems clear to us is that the asteroid has a first variation of the semi-major axis, through which it will pass through a specific region that may intensify recurrent perturbations due to the gravitational attraction with the Earth. Using equation 2, we will be able to analyze what is happening with the energy of the asteroid. We notice that it loses or gains energy with the variation of the semi-major axis. The variations of energy are enough for the orbit of the asteroid to change, making the long-term variations that we found and demonstrated in Figure 2.

\begin{figure*}[h!]
    \begin{center}
        \includegraphics[width=0.48\textwidth]{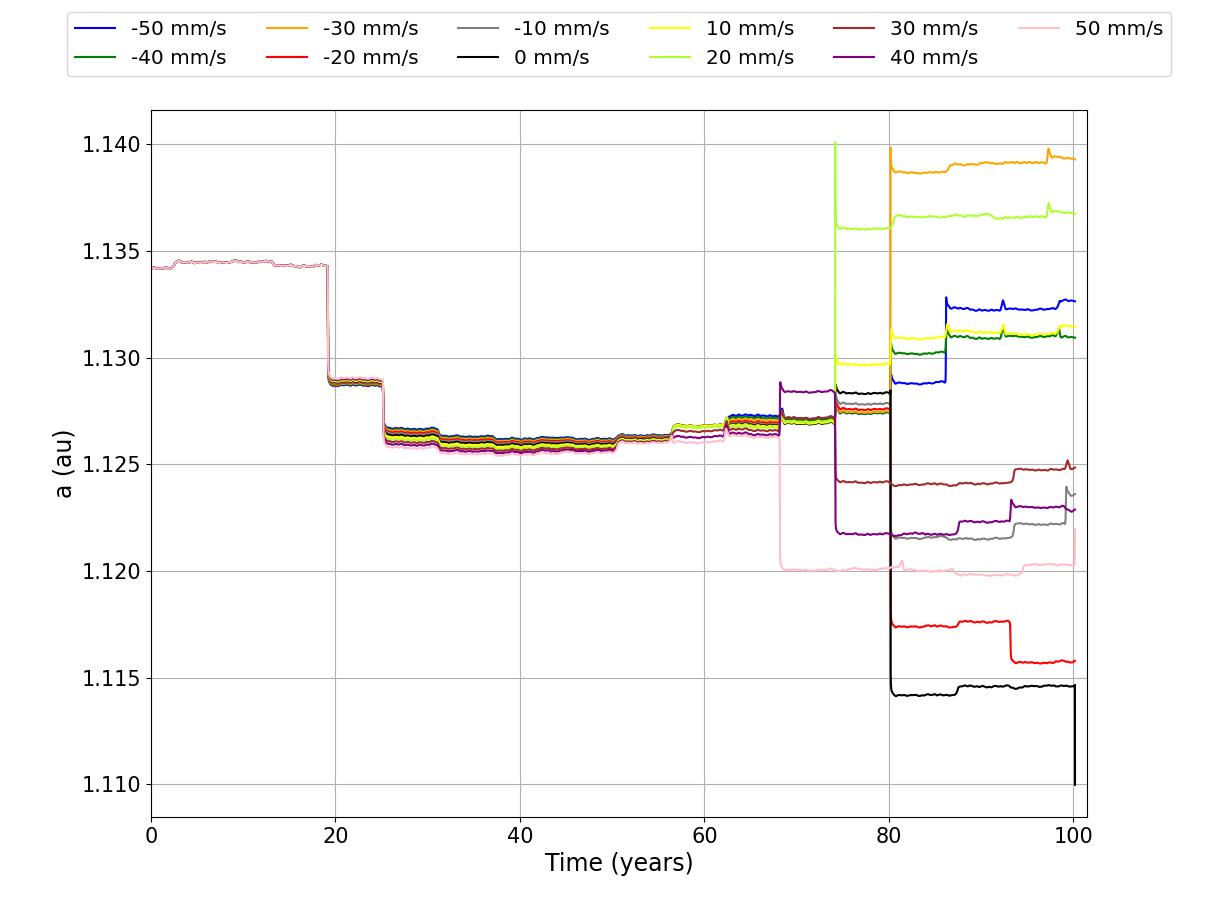}
        \quad
        \includegraphics[width=0.48\textwidth]{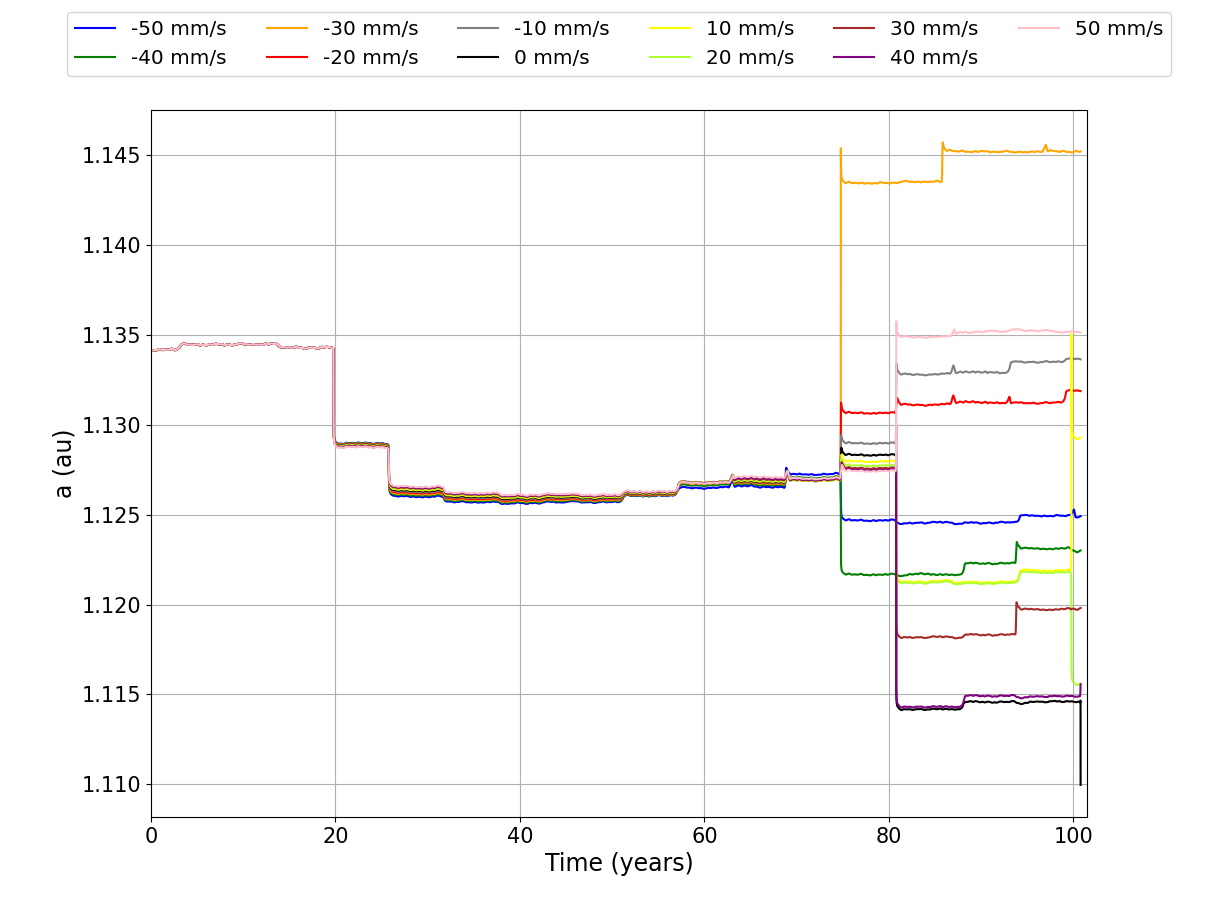}
        \caption{\small{Evolution of the semi-major axis of the asteroid as a function of the time after the impulses. Impulses applied at perihelion (left) and impulses applied at aphelion (right)}}
        \label{fig:fig1}
    \end{center}
\end{figure*}

As we can see in equations 4 and 5, the variations of the relative distances between the asteroid and the Earth is related to the gravity-assisted maneuver. We notice that, with the variations that occur with the approaches between the asteroid and the Earth, its energy vary, as we have already mentioned in the analysis of the semi-major axis. In other words, observing our results, we see that the main definer of the behavior of the orbit of the asteroid over a long period after the impulses are applied in specific regions, as we did here, is the gravitational perturbation coming from Earth in passages previous to the dangerous one. 

Figure 4 (left) shows the differences in the eccentricity of the orbit of the asteroid after the impulses at perihelion and at aphelion compared to the eccentricity without the impulses. We find that the eccentricity changes after the first close encounter and intensifies in the encounter that occurs between 70 and 80 years, suffering a significant variation after 80 years. These variations change the velocity of the asteroid, making it to reduce or increase velocity according to equation 3 and Figure 4 (right), which shows the variation of the velocity of the asteroid with the application of the impulse. It can be seen in Figure 4 (right) that the asteroid has a large variation in velocity with the impulse applied at perihelion, where, according to Figures 1 and 2, it was the impulse that moved the asteroid further away from Earth.

\begin{figure*}[h!]
    \begin{center}
        \includegraphics[width=0.48\textwidth]{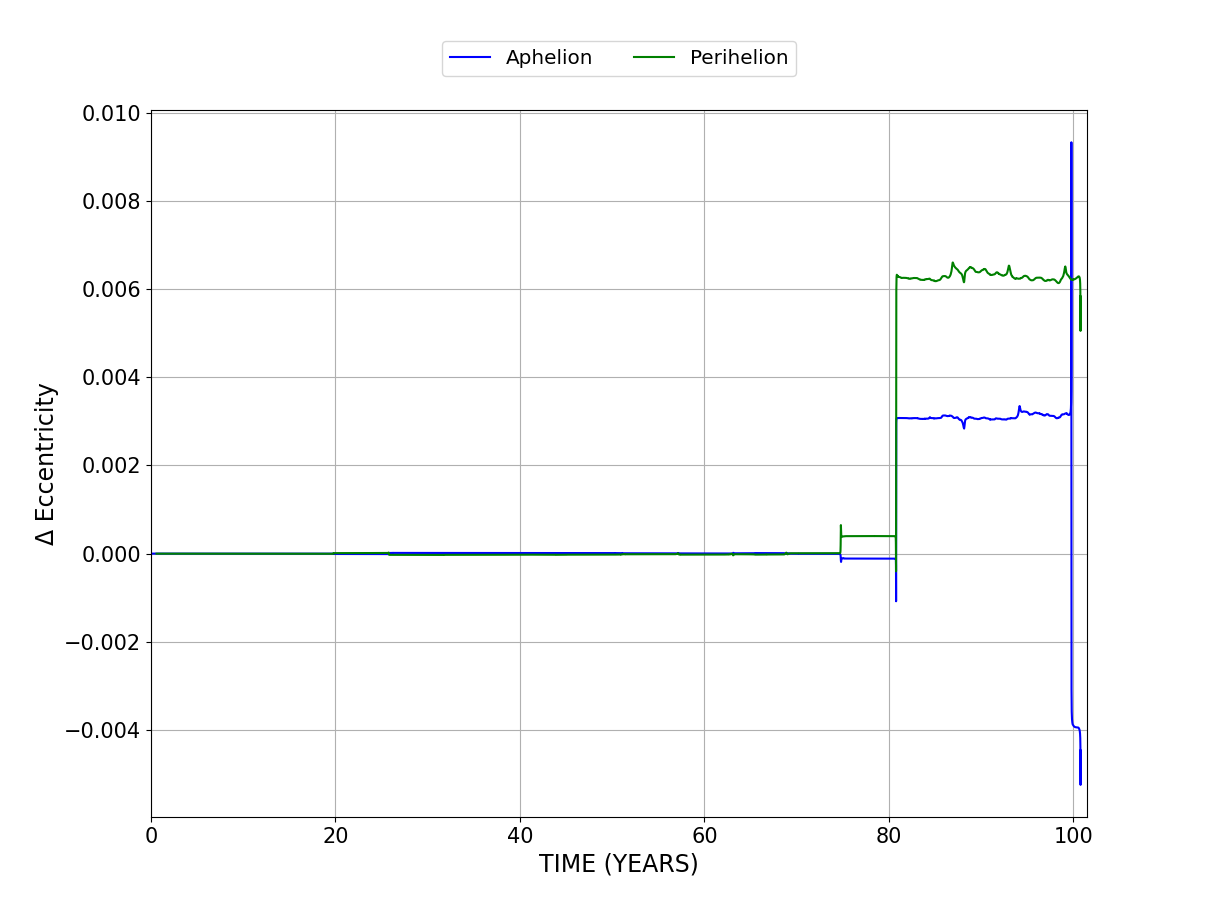}
        \quad
        \includegraphics[width=0.48\textwidth]{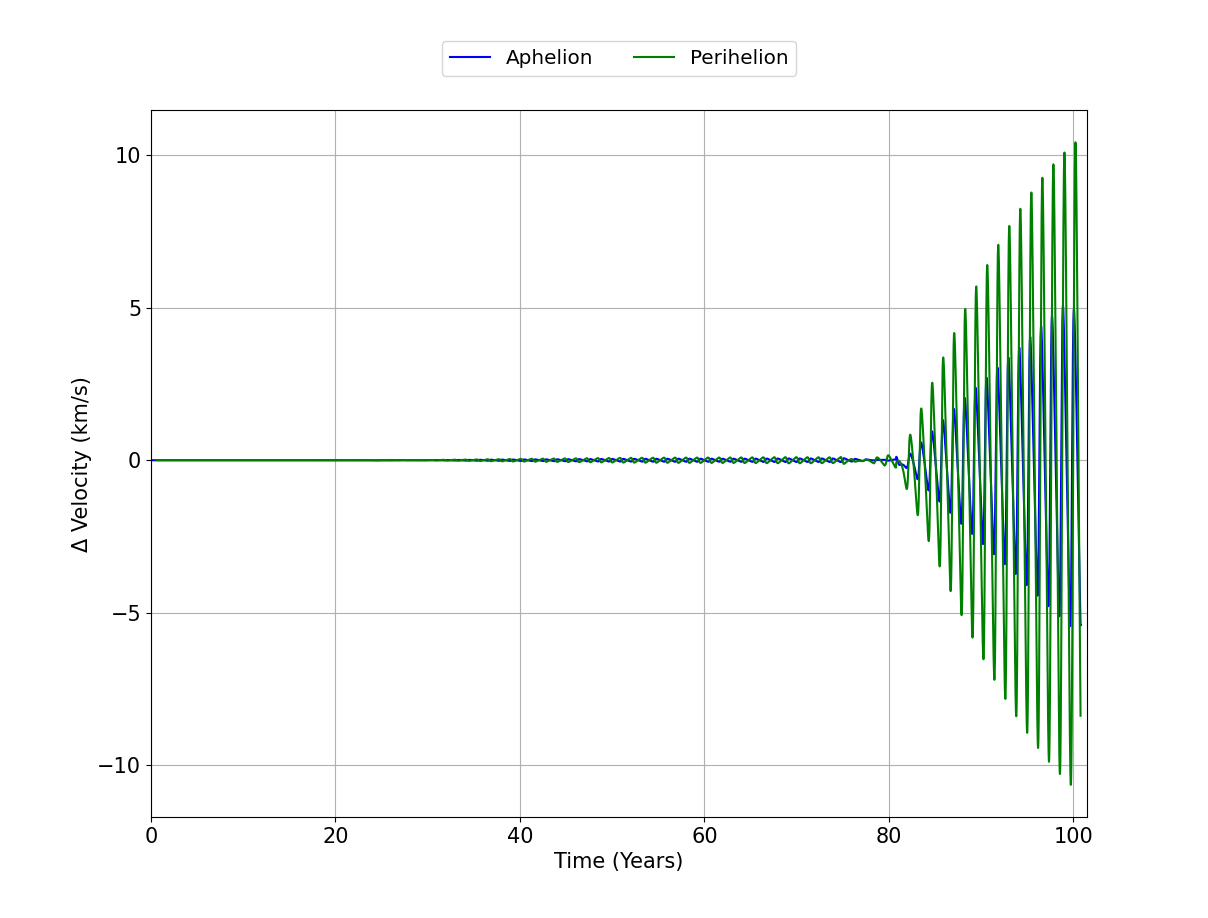}
        \caption{\small{Eccentricity differences (left) for the 10 mm/s impulse applied at perihelion (green) and aphelion (blue). Velocity differences (right) using the same impulse.}}
        \label{fig:fig1}
    \end{center}
\end{figure*}

Therefore, the variations of the semi-major axis and the orbital period of the asteroid are also modified. Therefore, the asteroid actually advance or delay in its next encounter with the Earth. These situations can be found in Figures 5 and 6.

Figure 5 shows the differences in the orbital period, in days, of the asteroid after the impulse is applied. Clearly the orbital period decreases when the larger close encounters occur between 70 and 85 years with the thrust applied at perihelion. We notice a small peak where the orbital period increases for the impulse applied at aphelion in this same time interval, but then the scenario of decreasing the orbital period of the asteroid occurs again.

\begin{figure*}[h]
    \begin{center}
        \includegraphics[width=0.95\textwidth, height=3.65in]{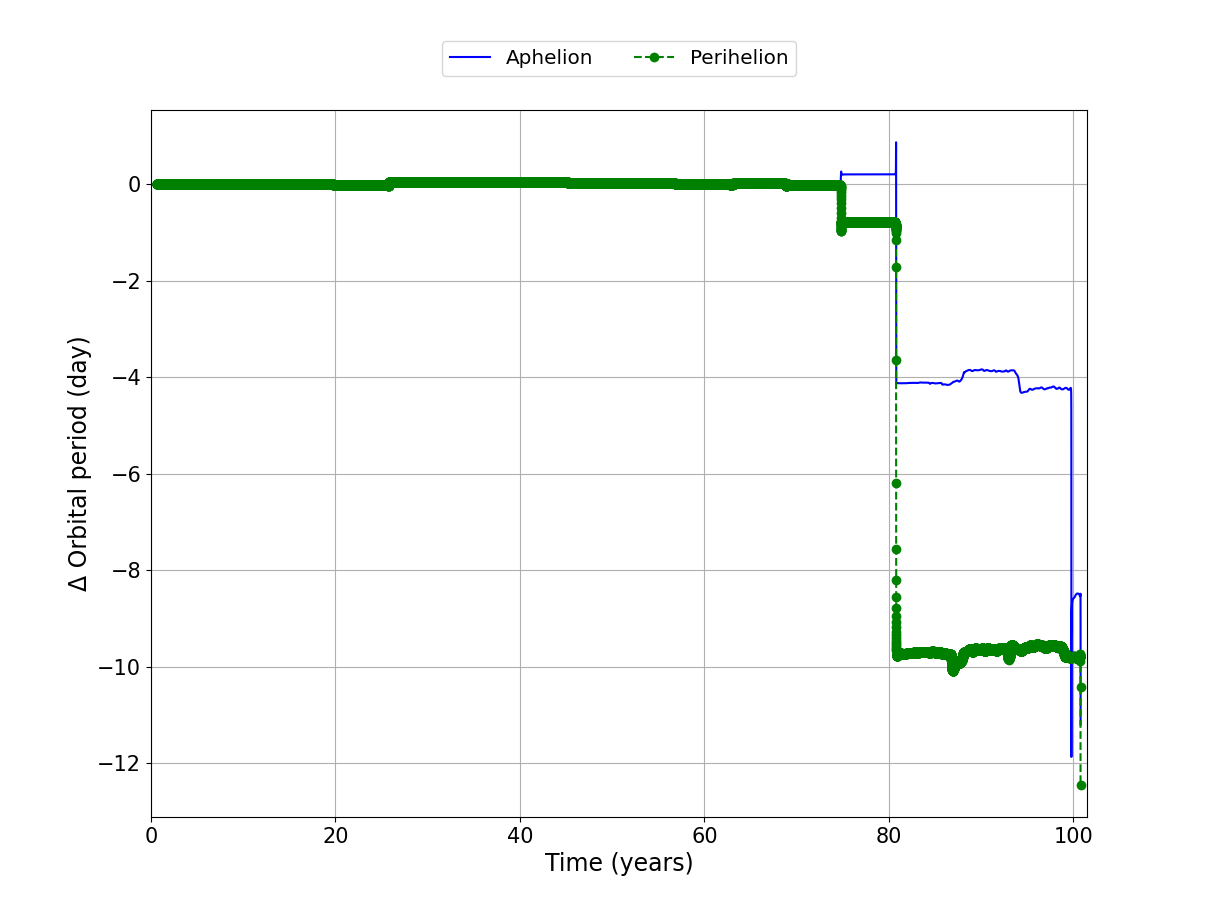}
        \caption{\small{Differences of the Orbital Period of the asteroid using the 10 mm/s impulse applied at perihelion (green) and aphelion (blue).}}
        \label{fig:fig1}
    \end{center}
\end{figure*}

\begin{figure*}[h]
    \begin{center}
        \includegraphics[width=0.95\textwidth, height=3.65in]{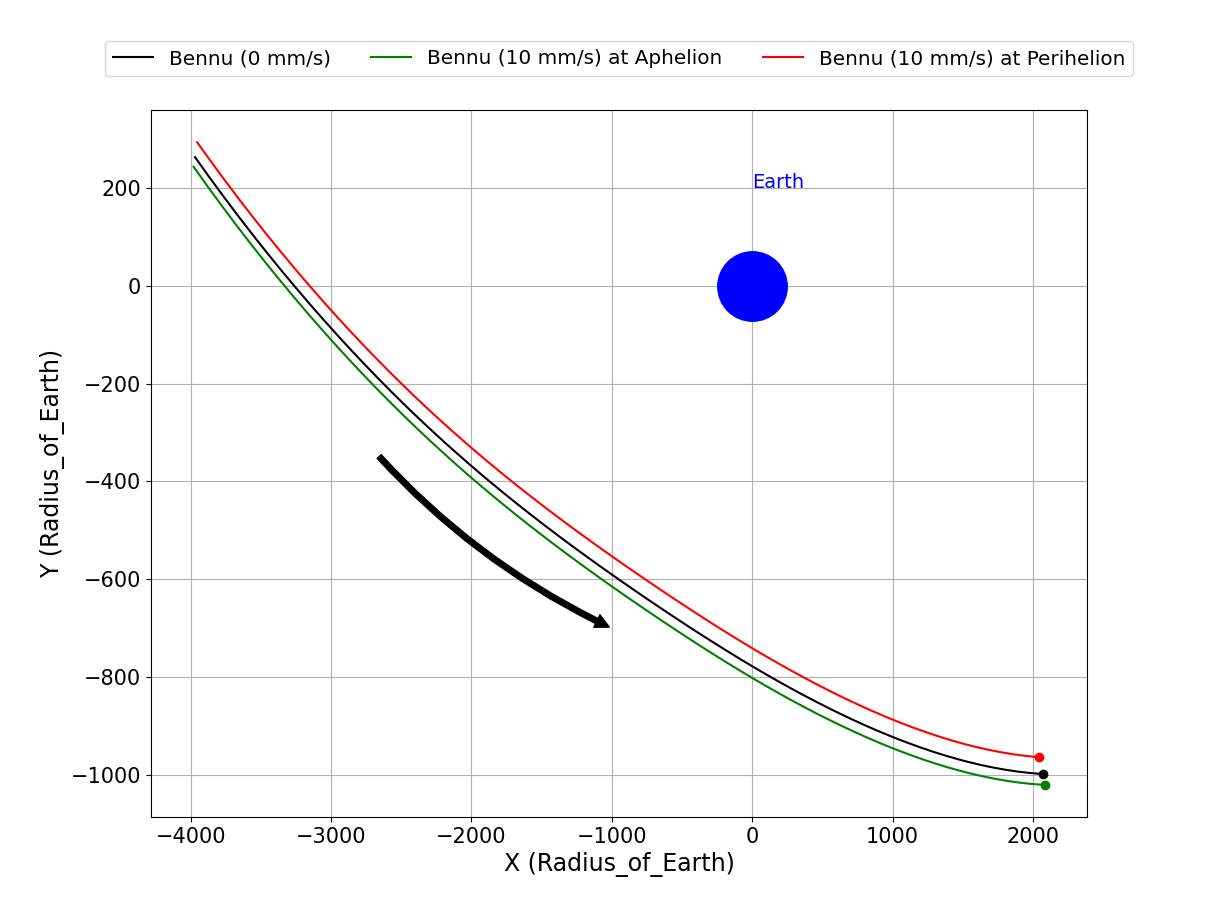}
        \caption{\small{Movement of the asteroid relative to Earth for the second close encounter.}}
        \label{fig:fig1}
    \end{center}
\end{figure*}

To demonstrate this delay or advance of the movement of the asteroid due to the variation of its orbital period, we can see in Figure 6 the moment of the second close encounter between the asteroid and the Earth. The blue dot representing the Earth is in the center of the reference system and the asteroid is performing its movement in an interval of 50 days. The arrow indicates the movement of the asteroid and the black, green and red lines indicate its movement with respect to the Earth and the dots at the end indicate the final position of the asteroid in this time interval. We can see from this Figure that, when applying the impulse at aphelion, the asteroid advances its motion and, on the other hand, when applying the impulse at the perihelion of the orbit of the asteroid, it is noticed that it is delayed in relation to the other points. This Figure quantifies what we explained earlier.

It is worth remembering that we studied here only the impulse of 10 mm/s, due to the results found in Figures 1 and 2. Therefore, all the results presented regarding the variations in eccentricity, velocity, orbital period and relative motion are related to this impulse. However, the explanations can be used for all other impulses. Everything is related to the region where the asteroid will pass due to the impulse applied and the variation of gravitational perturbations, where in some cases it will be enough to bring the asteroid even closer to the Earth and in others it moves it away.

\section{Conclusion}\label{sec:conclusion}

Our work shows the importance of considering long-term gravitational perturbations when developing asteroid deflection techniques. We found that it is not only the magnitude of the impulse that will be responsible for deflecting the asteroid, but rather the various close encounters that the asteroid has with the Earth. This is very encouraging, since we would not need a large impactor to cause a greater impulse, remembering that these close encounters will change as a new approach occurs due to the intensification of the gravitational perturbation, thus intensifying the swing-by maneuver.

We show that there is a variation in the orbital period of the asteroid and it can delay or advance its movement making it to arrive before the Earth in the regions where the close encounter would occur, or delay its movement, making the Earth to arrive first in the region of the closest encounter. Following this reasoning, it is also evident that the asteroid can be approached by the Earth due to the same phenomenon, that is, in regions where the asteroid and the Earth are at a considerable distance, the asteroid can advance or delay its movement and end up meeting with Earth.

We show that there is no generalization of this study. Each impulse generates a different result, as does each impulse application location.

\section*{Acknowledgements}
We would like to thank FAPESP proc. 2016/024561-0, 2018/17864-1 and CNPq proc 305210/2018-1, 309089/2021-2 for funding and contributing to this work. Coordination for the Improvement of Higher Education Personnel (CAPES) project PrInt CAPES-INPE.

\section*{Data availability}
The data underlying this article will be shared on reasonable request to the corresponding author.




\bibliographystyle{mnras}
\bibliography{references} 





\bsp	
\label{lastpage}
\end{document}